\documentclass[twocolumn,showpacs,prl,amsmath,amssymb,superscriptaddress,prl]{revtex4-1}

\usepackage[pdftex]{graphicx}
\usepackage{dcolumn}
\usepackage{bm}

\begin{document}

\preprint{APS/123-QED}

\title{
Fate of $\bm{yz}$/$\bm{zx}$ orbital degeneracy and $\bm{xy}$ Fermi surface in Ru substituted FeSe$_{\bm{1-x}}$Te$_{\bm{x}}$
}

\author{T. Sugimoto}
\affiliation{Department of Complexity Science and Engineering, University of Tokyo, 5-1-5 Kashiwanoha, Kashiwa 277-8561, Japan}

\author{D. Ootsuki}
\affiliation{Department of Physics, University of Tokyo, 5-1-5 Kashiwanoha, Kashiwa 277-8561, Japan}

\author{K. Sawada}
\affiliation{Department of Complexity Science and Engineering, University of Tokyo, 5-1-5 Kashiwanoha, Kashiwa 277-8561, Japan}

\author{H. Anzai}
\affiliation{Hiroshima Synchrotron Radiation Center, Hiroshima University, Higashihiroshima, Hiroshima 739-0046, Japan}

\author{M. Arita}
\affiliation{Hiroshima Synchrotron Radiation Center, Hiroshima University, Higashihiroshima, Hiroshima 739-0046, Japan}

\author{H. Namatame}
\affiliation{Hiroshima Synchrotron Radiation Center, Hiroshima University, Higashihiroshima, Hiroshima 739-0046, Japan}

\author{M. Taniguchi}
\affiliation{Hiroshima Synchrotron Radiation Center, Hiroshima University, Higashihiroshima, Hiroshima 739-0046, Japan}
\affiliation{Graduate School of Science, Hiroshima University, Higashihiroshima, Hiroshima 739-8526, Japan}

\author{M. Horio}
\affiliation{Department of Physics, University of Tokyo, 7-3-1 Hongo, Bunkyo-ku, Tokyo 113-0033, Japan}

\author{K. Horiba}
\affiliation{Institute of Materials Structure Science, High Energy Accelerator Research Organization (KEK), Tsukuba 305-0801, Japan}

\author{M. Kobayashi}
\affiliation{Institute of Materials Structure Science, High Energy Accelerator Research Organization (KEK), Tsukuba 305-0801, Japan}

\author{K. Ono}
\affiliation{Institute of Materials Structure Science, High Energy Accelerator Research Organization (KEK), Tsukuba 305-0801, Japan}

\author{H. Kumigashira}
\affiliation{Institute of Materials Structure Science, High Energy Accelerator Research Organization (KEK), Tsukuba 305-0801, Japan}

\author{T. Inabe}
\affiliation{Department of Applied Physics, Tohoku University, Sendai 980-8579, Japan}

\author{T. Noji}
\affiliation{Department of Applied Physics, Tohoku University, Sendai 980-8579, Japan}

\author{Y. Koike}
\affiliation{Department of Applied Physics, Tohoku University, Sendai 980-8579, Japan}

\author{N. L. Saini}
\affiliation{Dipartimento di Fisica, Universit\'a di Roma "La Sapienza", Piazzale Aldo Moro 2, Roma 00185, Italy}

\author{T. Mizokawa}
\affiliation{Department of Complexity Science and Engineering, University of Tokyo, 5-1-5 Kashiwanoha, Kashiwa 277-8561, Japan}
\affiliation{Department of Physics, University of Tokyo, 5-1-5 Kashiwanoha, Kashiwa 277-8561, Japan}
\affiliation{Dipartimento di Fisica, Universit\'a di Roma "La Sapienza", Piazzale Aldo Moro 2, Roma 00185, Italy}

\date{\today}

\begin{abstract}
We have investigated the impact of Ru substitution on the multi-band 
electronic structure of FeSe$_{1-x}$Te$_x$ by means of angle-resolved 
photoemission spectroscopy (ARPES).
The ARPES results exhibit suppression of the $xy$ Fermi surface
and the spectral broadening near the zone boundaries, which can be 
associated with the lattice disorder introduced by the Ru substitution.
The degeneracy of the Fe 3$d$ $yz$/$zx$ bands at the zone center, 
which is broken in FeSe$_{1-x}$Te$_x$, is partly recovered with the Ru substitution,
indicating coexistence of nematic and non-nematic electronic states.
\end{abstract}

\pacs{74.25.Jb, 74.70.Xa, 79.60.-i, 74.81.-g}
\maketitle

\newpage


The discovery of high-$T_c$ superconductivity in Fe pnictides \cite{Kamihara2008} 
has stimulated extensive experimental and theoretical investigations on 
the multi-orbital character of these. 
Among the Fe-based superconductors, FeSe$_{1-x}$Te$_x$ has the simplest 
crystal structure (anti-PbO type structure) with stacking of the FeSe(Te) layers
and is thought to be the most suitable for the study of multi-orbital physics 
\cite{Hsu2008}.
The electronic phase diagram of FeSe$_{1-x}$Te$_x$ is rich and very interesting
due to spin and lattice instabilities that are probably related to
the Fe 3$d$ multi-orbital character. 
One of the end members FeSe exhibits orthorhombic distortion below 90 K and superconductivity
below 8 K \cite{McQueen2009}. The other end member FeTe becomes antiferromagnetic
accompanied by orthorhombic lattice distortion \cite{Bao2009}.
FeSe$_{1-x}$Te$_x$ is basically tetragonal and exhibits superconductivity
with a maximum $T_c$ $\sim$ 15 K \cite{Noji2010}.
Moreover, the charge neutral cleavage plane is an extra advantage
to study multi-band Fermi surfaces of FeSe$_{1-x}$Te$_x$ 
using angle-resolved photoemission spectroscopy (ARPES).
In spite of the apparent simplicity of the crystal structure of FeSe$_{1-x}$Te$_x$,
the assignment of the orbital character in ARPES is still controversial.
For example, out of the four hole bands observed near the zone center in ARPES, 
the outermost hole band (which is labeled as $\gamma$ in several literatures) 
forms the large Fermi pocket and is always very weak. 
This outermost hole band is assigned to $xy$ by Chen {\it et al.} 
using polarization dependent ARPES \cite{Chen2010},
consistent with most of the ARPES studies \cite{Nakayama2010,Sudayama2013},
while Tamai {\it et al.} assign it to $yz/zx$ \cite{Tamai2010}.
The intermediate hole band (labeled as $\beta$ in literatures) 
forms the small Fermi pocket and is assigned to $yz/zx$ \cite{Chen2010} 
or $xy$ \cite{Okazaki2013}.
The inner most band that almost touches the Fermi level (labeled as $\alpha$ 
in literatures) is assigned to $yz/zx$ \cite{Chen2010,Okazaki2013}. 
In addition, Okazaki {\it et al.} \cite{Okazaki2013} 
resolve another intermediate hole band which does not reach 
the Fermi level and is almost degenerate with the inner most hole band 
at the zone center.  
We will call this intermediate hole band as $\beta'$.
The orbital degeneracy of $yz/zx$ bands ($\alpha$ and $\beta'$) 
at the $\Gamma$ point is likely to be confirmed by Okazaki {\it et al.} 
although it is not observed in most of the ARPES works.
Another possibility is that the hole band assigned to $xy$  by Okazaki {\it et al.} 
(corresponding to $\beta$) is one of the $yz/zx$ bands and, instead, the inner most band 
assigned to $yz/zx$ by Okazaki {\it et al.} (corresponding to $\alpha$) would be 
$xy$ as proposed by Sudayama {\it et al.} \cite{Sudayama2013}.
With this orbital assignment, the orbital degeneracy of $yz/zx$ is actually removed 
even in the result by Okazaki {\it et al.} in which the $\beta$ and $\beta'$ are 
successfully resolved. 
Such a puzzling situation indicates complexity of the electronic states in FeSe$_{1-x}$Te$_x$
that has been overlooked in the interpretation of ARPES results.

Local structural studies have revealed 
that the Se and Te atoms are distributed with short Fe-Se bonds and 
long Fe-Te bonds \cite{Joseph2010}. The different bond length and 
chalcogen height between Fe-Se and Fe-Te indicate that the ligand 
field splitting between $xy$ and $yz/zx$ can be reversed between
the elongated FeTe$_4$ and compressed FeSe$_4$ tetrahedron and that
the $yz/zx$ orbital degeneracy can be removed in the Fe(Se,Te)$_4$
tetrahedron. If the $yz/zx$ orbital degeneracy is locally broken, then
the band splitting between the $yz/zx$ bands at the $\Gamma$ point can be 
explained naturally. In addition, the crystal field reverse  
between $xy$ and $yz/zx$ can provide a possible explanation
of the two $xy$ (outer and inner) bands near $\Gamma$ point 
as pointed out by Sudayama {\it et al.} \cite{Sudayama2013}.
In this context, since the electronic structure of FeSe$_{1-x}$Te$_x$ 
is complicated by the local symmetry breaking, 
more systematic investigations are required to understand the intervening
coupling between the local disorder and the multi-orbital electronic states.

In order to address this issue, we focus on isovalent Ru substitution effect
on the multi-band electronic structure of FeSe$_{1-x}$Te$_x$. 
It is known that the transition-metal impurity can enhance anisotropic transport
in Fe pnictides \cite{Nakajima2011,Ishida2013}, 
and the origin of the anisotropy is assigned to anisotropic impurity scattering
\cite{Gastiasoro2014}. Therefore, it is highly important to study 
the transition-metal impurity effect on the multi-band electronic state 
by means of ARPES.
In this paper, we report an ARPES study on Ru substituted FeSe$_{1-x}$Te$_x$ 
which reveals the impact of local symmetry breaking on the multi-band 
electronic structure of FeSe$_{1-x}$Te$_x$. 

Single crystals of Fe$_{0.95}$Ru$_{0.05}$Se$_{0.3}$Te$_{0.7}$ were grown by the Bridgman method and annealed at 400 $^{\circ}$C for 200 h in vacuum (～10$^{-6}$ Torr) \cite{Noji2010,Inabe2013}.
The ARPES measurements at $h\nu$ = 23 eV were performed at beamline 9A, 
Hiroshima Synchrotron Radiation Center (HSRC) which has a normal 
incidence monochromator with off-plane Eagle mounting. 
Circularly polarized light from the helical undulator was exploited 
for the ARPES experiments.
The photoelectrons were collected using a SCIENTA R4000 analyzer. 
Total energy resolution including the monochromator and the electron 
analyzer was set to 19.9 meV.
The angular resolution was set to $\sim$ 0.2$^{\circ}$ that 
gives the momentum resolution of 0.0077 \AA$^{-1}$ for $h\nu$ = 23 eV.
The base pressure of the spectrometer was $10^{-9}$ Pa range. 
We cleaved the single crystals at 20 K under the ultrahigh vacuum 
and took ARPES data at the same temperature within four hours after the cleavage.
The ARPES measurements for $h\nu$ = 37 to 56 eV were performed 
at beamline 28A of Photon Factory, KEK using a SCIENTA SES-2002 
electron analyzer with circularly polarized light in order to avoid
the relatively strict selection rule of the linear polarization 
and to observe various Fe 3$d$ orbital symmetries.
The total energy resolution was set to 26.2 - 50.7 meV for $h\nu =$ 37 - 56 eV. 
The angular resolution was set to $\sim$ 0.2$^{\circ}$ that 
gives the momentum resolution of 0.010 - 0.013 \AA$^{-1}$ for $h\nu$ =37 - 56 eV.
The base pressure of the spectrometer was in the $10^{-9}$ Pa range. 
The spectra were acquired at 20 K within 12 hours after the cleaving.
In all the ARPES measurements, the single crystals were properly oriented 
on the sample stage by the standard Lau\'{e} measurements and were cooled 
using liquid He refrigerator. 
The Fermi level ($E_F$) was determined using the Fermi edge of gold reference samples.

\begin{figure}
\includegraphics[width=8.5cm]{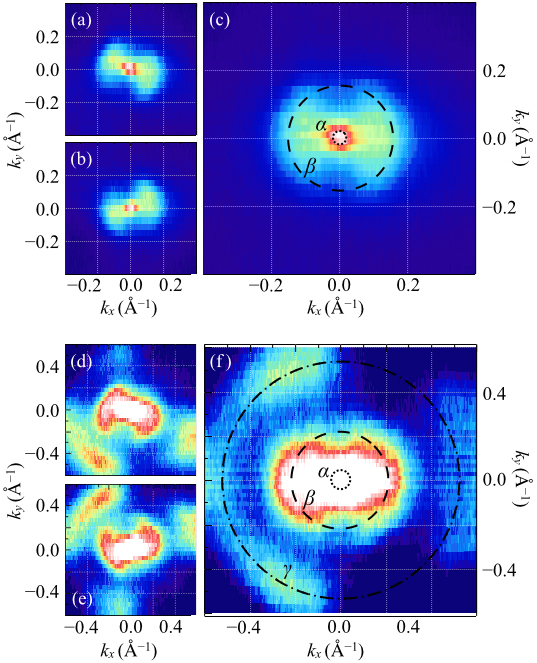}
\caption{(Color online) Fermi surface maps (integrated within $\pm$ 5 meV) with 
 (a) left-handed circular polarization (LCP), (b) right-handed circular polarization (RCP), 
and (c) the sum of the LCP and RCP. The dotted and dashed circles roughly indicate the position
of the $\alpha$ and $\beta$ Fermi surfaces. 
Photoemission intensity maps at 20 meV below $E_F$ (integrated within $\pm$ 5 meV)
with (d) LCP, (e) RCP, 
and (f) the sum of the LCP and RCP. The photon energy for all the maps is fixed at 23 eV ($k_z = 2.97$ \AA$^{-1}$, 
which is near the $\Gamma$ point). 
The dotted, dashed, and dot-dashed circles roughly indicate the position 
of the $\alpha$, $\beta$, and $\gamma$ bands. 
}
\label{fig1}
\end{figure}

Figures 1(a) and (b) show the photoemission intensity maps at $E_F$ with left- and right-handed 
circular polarization (LCP and RCP), respectively. 
The top of the $\alpha$ band (almost touching $E_F$) at the zone center 
and the $\beta$ Fermi surface are clearly observed. 
The butterfly-like shape of the $\beta$ Fermi surface,
which is induced by the transition-matrix element effect, is rotated clockwise
or counterclockwise depending on the helicity of the circularly polarized light.
This indicates that the $\beta$ Fermi surface is derived from 
one of the Fe $3d$ $yz$/$zx$ bands.
The intense $\alpha$ band at the zone center should be assigned to the other 
Fe $3d$ $yz$/$zx$ band since the suppression of the Fe $3d$ $xy$ band is expected
at the zone center due to the four-fold symmetry around the surface normal axis.
The sum of the LCP and RCP maps is displayed in Figure 1(c) where the two Fermi surfaces
are more clearly seen. 
Figures 1(d) and (e) show the photoemission intensity maps at 20 meV below $E_F$ 
with LCP and RCP, respectively. 
In addition to the $\alpha$ and $\beta$ bands seen in the Fermi surface maps,
the outermost $\gamma$ band is clearly observed. The sum of the LCP and RCP maps 
is displayed in Figure 1(f) and the three hole bands are more clearly seen.
The absence of $\gamma$ Fermi surface and the existence of $\gamma$ band at 20 meV
suggests gap opening at $E_F$ in the $\gamma$ band.

\begin{figure}
\includegraphics[width=8.5cm]{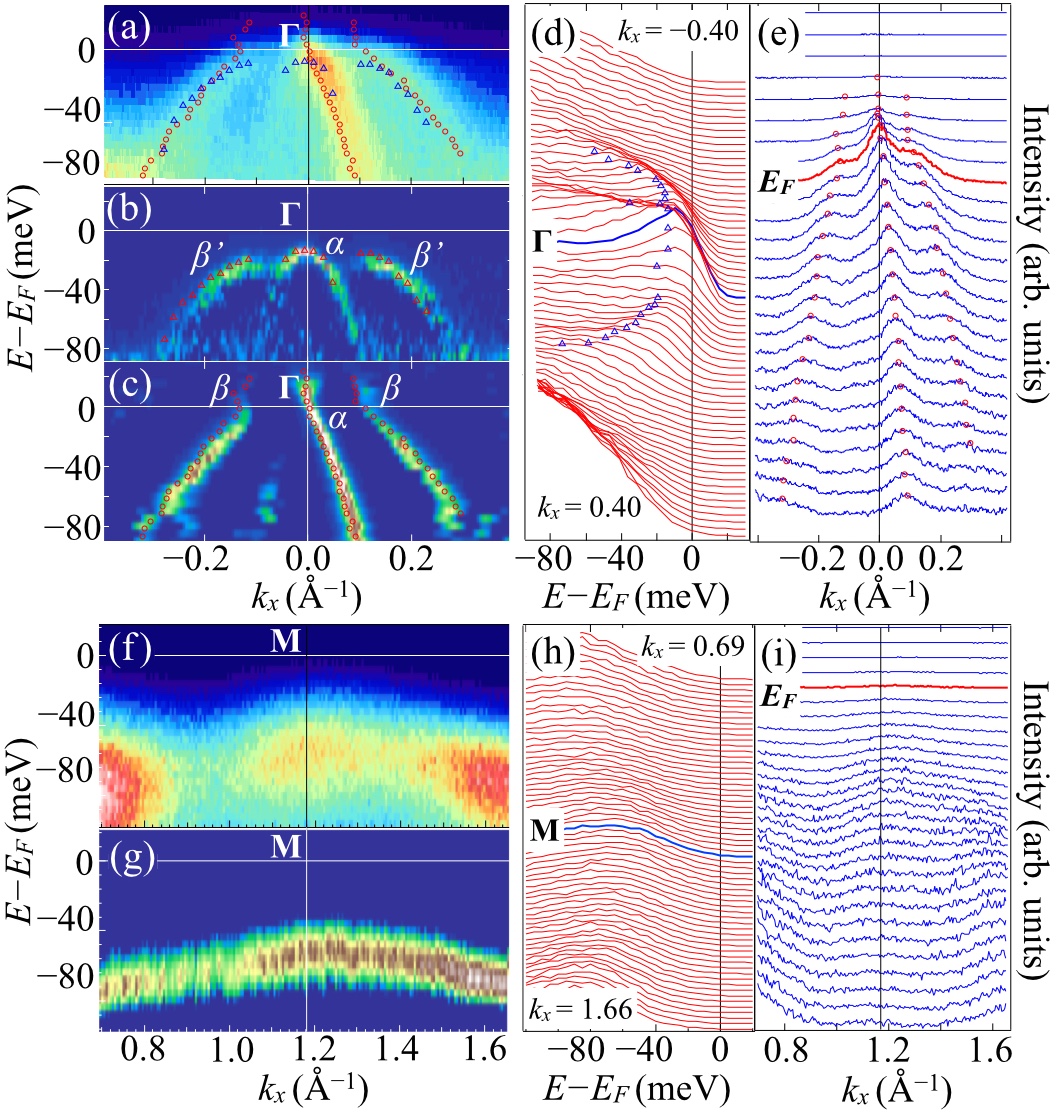}
\caption{(Color online)  (a) Cuts around the $\Gamma$ point along $k_x$,
(b) curvature of EDCs, (c) curvature of MDCs (d) EDCs, and (e) MDCs. (f) Cuts around M point along $k_x$ ,
 (g) curvature of EDCs, (h) EDCs, and (i) MDCs.
The circles and triangles indicate the band positions obtained from
the MDC and EDC, respectively.}
\label{fig2}
\end{figure}

Figures 2(a), (b), and (c) show the band dispersions along the cut 
around the $\Gamma$ point as a function of $k_x$.
Obtained from the energy distribution curves (EDCs) in Figure 2(d), the band dispersion labeled as $\beta'$ as well as from their curvatures \cite{Zhang2011} in Figure 2(b) does not reach $E_F$ and
takes its maximum at the zone center. By contrast, the band dispersion of the $\beta$ band clearly crosses $E_F$, which is extracted from the momentum distribution curves (MDCs) in Figure 2(e). The same behavior can be seen in their curvatures in Figure 2(c).

This observation suggests that,
in addition to the $\beta$ band forming the $\beta$ Fermi surface, 
another hole like band, which can be labeled as $\beta'$, exists
around the zone center.
Indeed, as shown in Figures 2(d) and (e), the intensity of the $\beta$ band drops 
for $|k_x|$ smaller than $\sim$ 0.175 \AA$^{-1}$ in the EDC and MDC plots, 
indicating that the $\beta$ band crosses $E_F$ at $|k_x|\sim$ 0.175 \AA$^{-1}$ 
which corresponds to the $\beta$ Fermi surface in Figure 1. 
On the other hand, some intensity remains even for $|k_x|$ 
smaller than $\sim$ 0.175 \AA$^{-1}$ and reaches the zone center, where
this remaining band ($\beta'$) and the $\alpha$ band are almost degenerate. 

Since the tops of the $\alpha$ and $\beta'$ band dispersions 
derived from the $yz/zx$ bands are almost degenerate at the zone center,
the orbital degeneracy of $yz/zx$ remains at the zone center.
The remaining $yz/zx$ orbital degeneracy may contradict with the most of 
the published results including a recent ARPES work by Miao {\it et al.}
\cite{Miao2014}, that reported a ubiquitous $yz/zx$ splitting.
On the other hand, this degeneracy is consistent with the ARPES work by
Okazaki {\it et al.} \cite{Okazaki2013}.
In the present ARPES, the four hole bands
($\alpha$, $\beta$, $\beta'$, and $\gamma$) can be identified.
Since only the three hole bands are expected near $E_F$ in the homogeneous
FeSe or FeTe, the four hole bands indicate an electronic inhomogeneity to some extent.
The four hole bands can be interpreted as the superposition of the two band 
structures with and without the degeneracy of the $yz/zx$ bands at the zone center. 
Namely, the system exhibits a sort of phase separation 
into the two electronic states with and without the electronic nematicity. 
The state with $\alpha$, $\beta$, and $\gamma$ bands is likely to be accompanied 
by the static or dynamic orthorhombic distortion for breaking the $yz/zx$ degeneracy. 
This state could be similar to the anisotropic or nematic state
proposed by Miao {\it et al.} \cite{Miao2014}. 
The other state with $\alpha$, $\beta'$, and $\gamma$ bands keeps 
the tetragonal symmetry with the $yz/zx$ degeneracy and corresponds 
to the non-nematic state.
However, since the $\beta$ and $\beta'$ are not clearly separated,
the energy splitting between the $yz$ and $zx$ bands should have distribution
due to static inhomogeneity and/or dynamic fluctuation.
That is to say, the nematic state and the non-nematic state coexist in the Ru substituted system.

Without the Ru substitution, the $\beta'$ band is not clearly
observed under the same experimental condition \cite{Sudayama2013}. 
In addition, the intensity and energy of the $\alpha$ band is
strongly modified from the prediction of the band-structure calculation. 
In FeSe$_{1-x}$Te$_x$, the $\alpha$ band is located well below $E_F$ 
and is very broad even at the zone center.
Therefore, Sudayama {\it et al.} assigned the modified $\alpha$ band 
to the $xy$ band from the FeSe-like orbital state \cite{Sudayama2013}.
This position of the $xy$ band is actually consistent with the recent ARPES
study on FeSe \cite{Maletz2014}, partially supporting the interpretation, i.e.,
 it is possible that the $yz/zx$ orbital degeneracy is broken 
due to the participation of the $xy$ orbital in FeSe$_{1-x}$Te$_x$.
On the other hand, the present results on the Ru substituted system indicate that 
the modified $\alpha$ band in FeSe$_{1-x}$Te$_x$ becomes normal (viz, 
comes close to the prediction of the band-structure calculation 
for the average structure) and the $\beta'$ band appears just below $E_F$.
Consequently, the $yz/zx$ degeneracy ($\alpha$ and $\beta'$) 
at the zone center is recovered with the Ru substitution.
Interestingly, such behavior is observed in as-grown FeSe$_{1-x}$Te$_x$
without annealing treatment \cite{Okazaki2013,Sudayama},
suggesting that the static inhomogeneity is introduced by the Ru substitution 
or by the atomic disorder and the $yz/zx$ orbital degeneracy is partially recovered.

The band dispersions as functions of $k_x$ around the zone corner are displayed
in Figures 2(f) and (g), and the corresponding EDCs and MDCs are plotted in 
Figures 2(h) and (i). The EDC and MDC plots show that the spectral features 
are very broad near the zone corner compared to those of annealed FeSe$_{1-x}$Te$_x$
\cite{Sudayama2013}. On the other hand, the spectral features near 
the zone center are comparable to those of FeSe$_{1-x}$Te$_x$.
The similar momentum dependent broadening is observed in as-grown FeSe$_{1-x}$Te$_x$
without annealing treatment \cite{Sudayama}.
Such momentum dependent broadening suggests that the Ru substitution can 
introduce additional atomic disorder which can freeze the orbital/lattice fluctuations and reduce the $yz$/$zx$ splitting.

\begin{figure}
\includegraphics[width=5.5cm]{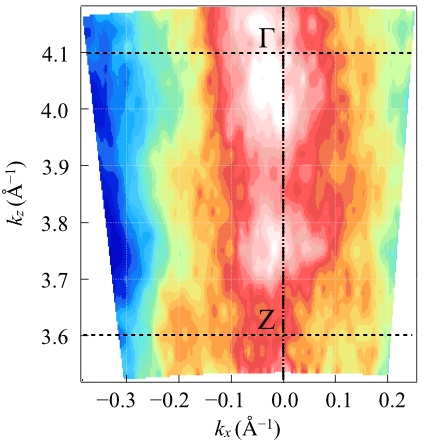}
\caption{(Color online) Fermi surface map in the $k_x$-$k_z$ plane.
The ARPES spectra are integrated within $\pm$ 18 meV relative to $E_F$.}
\label{fig3}
\end{figure}

Figure 3 shows a Fermi surface map in the $k_x$-$k_z$ plane taken 
at $h\nu$ = 37 to 56 eV, showing good two-dimensionality of the $\beta$ Fermi surface
although the intersection of the $\beta$ Fermi surface slightly increases in going 
from $\Gamma$ ($k_z = 0$) to Z ($k_z = \pi/c$).
The band-structure calculations with the average structure predict $k_z$ dependence
of the Fermi surfaces and the $yz$/$zx$ splitting. 
For example, the doubly degenerate $yz$/$zx$ bands are located below the $xy$ band 
at $\Gamma$ while they are located above the $xy$ band at Z in the calculations
\cite{Subedi2008, Miyake2010, Chen2010}.
The increase of the $\beta$ Fermi surface area with $k_z$ is in qualitative agreement
with the calculations. However, the area of the $\gamma$ Fermi surface is much larger
than the calculations.
The disagreement between the ARPES results and the calculations
can be assigned to the moderate correlation effect and 
the random distribution of Se and Te, that can strongly 
disturb the momentum dependent interlayer coupling and provide 
the good two dimensionality to the Fermi surfaces.

In Figure 4, the band dispersions, EDCs and MDCs around the $\Gamma$ and Z points of Figure 3 are displayed. 
The band dispersions around  $\Gamma$ are more or less consistent with those 
in Figure 2.
The $\alpha$, $\beta$, and $\beta'$ bands can be identified since
the intensity of the $\beta$/$\beta'$ band drops for $|k_x|$ smaller 
than $\sim$ 0.175 \AA$^{-1}$ but some intensity remains even for $|k_x|$ 
smaller than $\sim$ 0.175 \AA$^{-1}$ and reaches the zone center.
Again, this is consistent with the picture of the strong lattice disorder 
and the coexistence of the nematic and non-nematic electronic states.

\begin{figure}
\includegraphics[width=8.5cm]{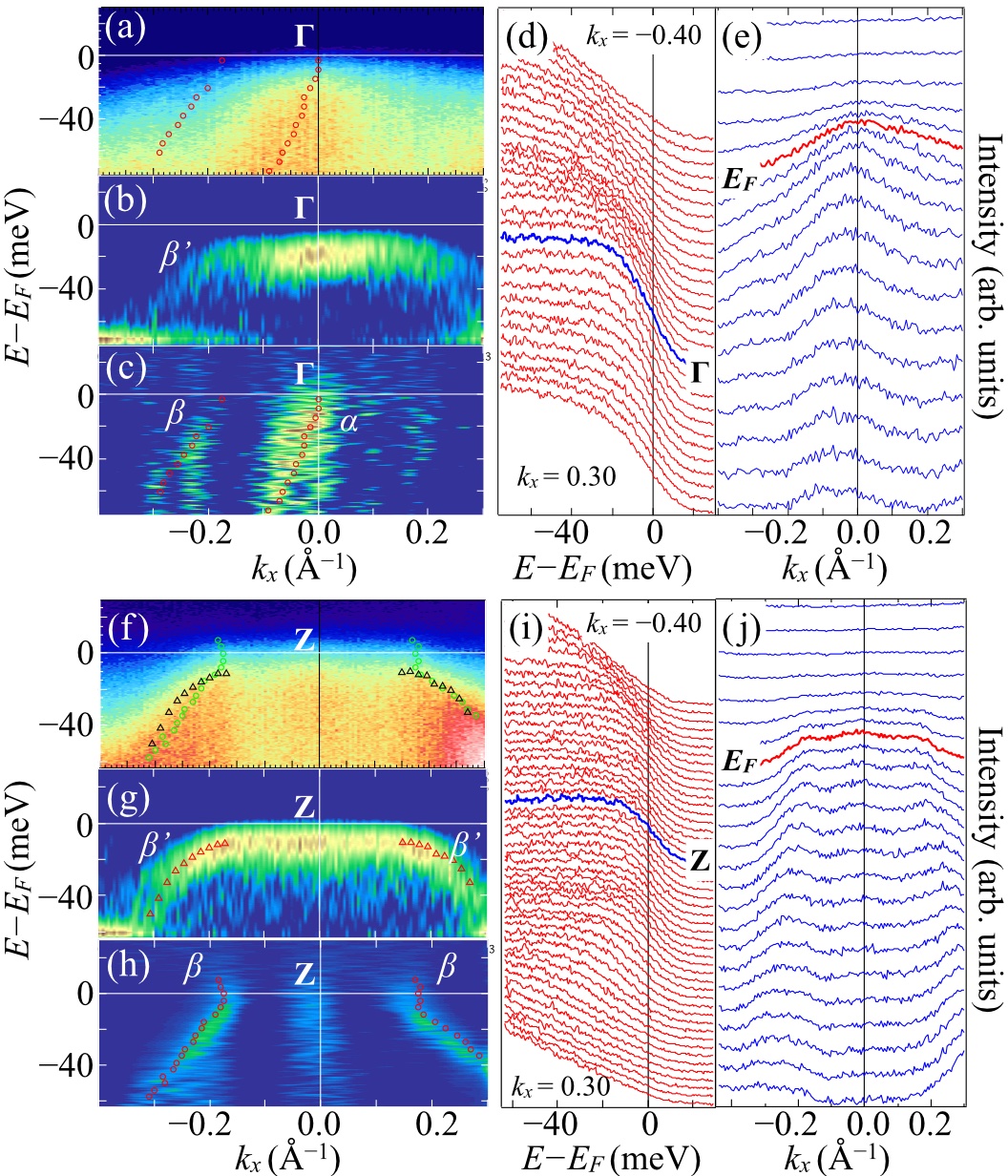}
\caption{(Color online)  (a) Cuts around $\Gamma$ point along $k_x$, 
 (b) curvature of EDCs, (c) curvature of MDCs, (d) EDCs, and (e) MDCs. (f) Cuts around Z point  along $k_x$, 
 (g) curvature of EDCs, (h) curvature of MDCs, (i) EDCs, and (j) MDCs. 
The circles and triangles indicate the band positions obtained from
the MDC and EDC, respectively.
}
\label{fig4}
\end{figure}

In summary, the multi-band electronic structure of the Ru 
substituted FeSe$_{1-x}$Te$_x$ has been studied by means of ARPES.
With the Ru substitution, the $\gamma$ Fermi surface with the Fe 3$d$ $xy$ 
character is suppressed and the band dispersions near the zone boundaries
are broadened. These observations indicate strong lattice disorder 
introduced by the Ru substitution.
While the degeneracy of the Fe 3$d$ $yz$/$zx$ bands at the zone center
is broken in FeSe$_{1-x}$Te$_x$, it is partly recovered in the Ru 
substituted FeSe$_{1-x}$Te$_x$. This behavior is further confirmed 
by the $k_z$ dependent ARPES measurements and shows that
the strong lattice disorder by the Ru substitution causes 
the coexistence of the nematic and non-nematic electronic states.
The moderate electronic correlation effect and the random distribution 
of Se and Te, which would play essential roles for the electronic 
nematicity in FeSe$_{1-x}$Te$_x$, are disturbed by the Ru substitution.

This work was partially supported by Grants-in-Aid from the Japan Society of 
the Promotion of Science (JSPS) (Grant No: 25400356).
T.S. and D.O. acknowledge supports from the JSPS Research Fellowship 
for Young Scientists.
The synchrotron radiation experiments have been done with the approval of 
Photon Factory, KEK (2013G021) and Hiroshima Synchrotron Radiation Center 
(Proposal No.13-A-6).

\end{document}